\documentclass[
nopreprint    													%choose either nopreprint or preprint 
]{jasatex}

\usepackage{amsmath,amsfonts}    % need for subequations
\usepackage{graphicx}   % need for figures
\usepackage{color}      % use if color is used in text
\usepackage{float}
\usepackage{fullpage}
\usepackage{bm}
\usepackage{subfig}
\DeclareGraphicsExtensions{.eps,.png,.pdf}

\renewcommand{\appendix}{
  \setcounter{section}{0}\renewcommand{\thesection}{\Alph{section}}
  \section*{Appendix} }

\def\eeq{\relax}
\def\beq#1#2\eeq{\begin{equation}\label{#1}#2\end{equation}}
\def\bal#1#2\eal{\begin{align}\label{#1}#2\end{align}}
\def\bse#1#2\ese{\begin{subequations}\label{#1}#2\end{subequations}}
\newcommand{\ca}{\begin{cases}}
\newcommand{\ac}{\end{cases}}
\newcommand{\ma}{\begin{pmatrix}}
\newcommand{\am}{\end{pmatrix}}

\newcommand{\er}\eqref

\begin{document}

\title{Tunable cylindrical shell as an element in acoustic metamaterial}

\author{Alexey S. Titovich}
\email{alexey.titovich@rutgers.edu}
\author{Andrew N. Norris}
%\email{norris@rutgers.edu}
%\homepage{http://www.rci.rutgers.edu/~norris/}
\affiliation{Mechanical and Aerospace Engineering, Rutgers University, Piscataway, NJ 08854}
%%%%%%%%%


%4340Ey Vibrations of shells
%4340Fz Acoustic scattering by elastic structures
%4320Fn Scattering of acoustic waves 
%4335Gk Phonons in crystal lattices, quantum acoustics 


\begin{abstract}
Elastic cylindrical shells are fitted with an internal mechanism which is optimized so that, in the quasi-static regime, the combined system exhibits prescribed effective acoustic properties. The mechanism consists of a central mass supported by an axisymmetric distribution of elastic stiffeners. By appropriate selection of the mass and stiffness of the internal mechanism, the shell’s effective acoustic properties (bulk modulus and density) can be tuned as desired. Subsonic flexural waves excited in the shell by the attachment of stiffeners are suppressed by including a sufficiently large number of such stiffeners. Effectiveness of the proposed metamaterial is demonstrated by matching the properties of a thin aluminum shell with a polymer insert to those of water. The scattering cross section in water is nearly zero over a broad range of frequencies at the lower end of the spectrum. By arranging the tuned shells in an array the resulting acoustic metamaterial is capable of steering waves. As an example, a cylindrical-to-plane wave lens is designed by varying the bulk modulus in the array according to the conformal mapping of a unit circle to a square.

\end{abstract}

\maketitle

\section{Introduction}   \label{sec1}
Transformation based design of materials for wave steering originated in electromagnetics. Using singular transformations and the invariance of Maxwell's equations under such transformations the possibility of wave steering was demonstrated~\cite{Pendry06}. The idea is to steer an incident wave in a finite region surrounding the object such that there is no scattering. Although the theory is frequency independent, only cloaking of objects at microwave frequencies has been achieved experimentally~\cite{Schurig06}. Cloaking objects at frequencies of visible light is  theoretically possible using conformal mapping as shown by Leonhardt~\cite{Leonhardt06}, but is experimentally unconfirmed. Other applications of transformation optics include beam shifters and splitters~\cite{Rahm08}.

By natural progression the ideas found applications in acoustics. Cummer et al.~\cite{Cummer07} showed the equivalent invariance under coordinate transformations between Maxwell's and Helmholtz' equations. Using singular transformations they designed a 2D acoustic cloak with anisotropic density and bulk modulus. Simulations showed that the cloak could steer waves around the annulus. Chen and Chan~\cite{Chen07} applied the same concept to the design of a 3D acoustic cloak. Cloaking of elastodynamic waves was also investigated~\cite{Milton06}.
One  way to achieve the anisotropic density required for inertial cloaking is by use of  layered fluids~\cite{Torrent08b, Norris10a}. However, such cloaking devices are not viable because the density requirement results in an infinitely massive cloak~\cite{Norris08b}. Unlike cylindrical acoustic cloaks, carpet cloaks have been experimentally shown to hide objects resting on a surface~\cite{Popa11}. Yet another application of transformation acoustics is a cylindrical-to-plane wave lens as designed by Layman et al.~\cite{Layman11}. It works by steering waves, due to a monopole source at the center, from the corners to the faces of the lens. Their device is based on constructive multiple scattering from finite embedded elastic materials in a fluid matrix, something previously investigated by Torrent and Sanchez-Dehesa~\cite{Torrent07}.

This article expands the possibilities in Ref.~\cite{Torrent07} demonstrating that it is possible to tune an elastic component in a fluid to yield desired acoustic properties at low frequencies. The component is an elastic shell fitted with an internal springs-mass system (oscillator). The oscillator is comprised of a central rod with $J$ identical and axisymmetrically distributed lengthwise ribs fixing it to the shell. The rod behaves as an added mass and the ribs provide additional stiffness. These two parameters together with the shell thickness determine the effective acoustic properties of the complete system. 

Acoustic scattering from an empty shell as studied by Bleich and Baron \cite{Bleich54} is very different from that for a stiffened shell. The addition of an internal mechanical oscillator to a shell was originally investigated by Achenbach et al.~\cite{Achenbach92}. They showed that a single spring-mass system excites flexural modes which are subsonic in an empty shell. Guo \cite{Guo92} presented the analysis for a diametrical pair of springs supporting a central mass. The geometry of the forcing results in a normal mode scattering solution which is of two-type for even and odd modes. Recently, Titovich and Norris \cite{Titovich13} obtained the scattering response for an axisymmetric distribution of an arbitrary number of springs supporting a central mass. Their results are applied here to demonstrate that it is possible to tune an elastic shell such that: (i) the density and speed of the shell-springs-mass system is matched to water and (ii) the flexural resonances are suppressed at low frequencies. 

The paper is organized as follows. The analytical solution for a shell with a central mass attached by $J$ axisymmetric linear springs is summarized in \S \ref{sec2}. The quasi-static acoustical properties of the system are then determined in \S \ref{sec3}. Material selection for the shell is described in \S \ref{sec4} as well as the presence of flexural resonances at low frequencies. As an example, an aluminum shell is tuned to water in \S \ref{sec4}. Two physical oscillator designs are proposed in \S \ref{sec5}. The effect of surrounding fluid for a shell-springs-mass system in a fluid saturated array is described in \S \ref{sec6}. Wave steering by such an array is demonstrated in \S \ref{sec6} with a cylindrical-to-plane wave lens constructed from a 7x7 array of tuned shells. Conclusions are presented in \S \ref{concl}.

\section{Scattering from a shell with an internal mass} \label{sec2}

We review the main results from Ref.~\cite{Titovich13} for later use. Consider an infinitely long thin elastic shell of external radius $a$, thickness $h \ll a$, density $\rho_s$, elastic modulus $E_s$, Poisson's ratio $\nu_s$ and extensional wave speed $c_p$ defined by  $c_p^2 = {E_s}/[{\rho_s (1-\nu_s^2)}]$. The internal mechanism consists of a central  mass $m$ per unit axial length supported by $J$ massless springs of  individual spring stiffness $\kappa$ (with units of stress). The mass is centered so that the springs act concentrically under small amplitude motion of the mass (see derivation in appendix of Ref.~\cite{Titovich13}).
 
The shell is surrounded by an acoustic fluid (water) of density $\rho$ and wave speed $c$ with bulk modulus $K=\rho c^2$. Time harmonic solutions are sought with factor $e^{-i\omega t}$ omitted as understood. 

The total acoustic pressure $p$ in the fluid satisfies  Helmholtz' equation $\nabla^2 p + k^2 p = 0$, and is decomposed into incident and scattered pressure fields as $p = p_i + p_s$, where $p_i = \sum\limits_{n = - \infty}^\infty  A_n J_n(kr)  \text{e}^{\text{i} n \theta}$ and $p_s = \sum\limits_{n = - \infty}^\infty  B_n H_n^{(1)}(kr) \text{e}^{\text{i} n \theta}, \quad r\ge a$, respectively, with  $H_n^{(1)}$  the Hankel function of the first kind of order $n$.
The scattered pressure  is expressed 
via  $\bf B=\bf T \bf A$ where $\bf B$ and $\bf A$ are infinite vectors with elements  $B_n$ and $A_n$, respectively, and $\bf T$ is the so-called  T-matrix.

The main result from Ref.~\cite{Titovich13} is that the T-matrix splits into the superposition of  the T-matrix for the empty shell plus  a $J$-cyclical contribution due to the springs-mass system
\begin{equation} \label{Tmatrix}
{\bf T} = {\bf T}^{(0)} +  \sum\limits_{j=1}^{J} {\bf b}_j {\bf b}_j^T  ,
\end{equation}
where ${\bf T}^{(0)} = \text{diag}(T_n)$ is the empty shell T-matrix   with components $T_n = ( {\zeta_n^*}-{\zeta_n}) /(2\zeta_n)$,  $\zeta_n = (Z_n^{sh} + Z_n) H_n^{(1)\prime}(ka)$. The acoustic impedance is   $Z_n = \text{i} \rho c {H_n^{(1)}(ka)}$$/{H_n^{(1)\prime}(ka)} $ and the shell impedance is $Z_n^{sh} = -\text{i} \rho_s c_p \frac{ h}{a} \Big[ \Omega- \frac{\beta^2 n^4}{\Omega} - \Big( \Omega- \frac{n^2}{\Omega} \Big)^{-1}\Big] $, where $\beta = \frac{1}{\sqrt{12}}\frac{h}{a}$ and the dimensionless frequency $\Omega$ is  $\Omega = {\omega a}/{c_p}$. The  other part of the T-matrix in eq.\ \eqref{Tmatrix} depends on $b_{j,n} = \frac{\text{i}}{\zeta_n}  \left(  \frac{2\rho c Z^{tot}_n}{\pi ka}\right)^{1/2}  \ \ \text{if}\ n=j\,\text{mod}\, J  , \  \text{otherwise}\ 0 $, where the resonant behavior of the shell-springs-mass system is governed by the total equivalent impedance $\frac 1{Z_n^{tot}} = \frac{1}{Z^{sp}_{n}}  +\sum\limits_{p = - \infty}^\infty \frac{1}{Z^{sh}_{n+pJ}+Z_{n+pJ}} $ in terms of  the spring-mass impedance $Z^{sp}_n$ and resonant frequency $\omega_{sp}$, 
\begin{equation} \label{Zsp}
\begin{aligned}
Z^{sp}_n(J) &= \frac{\text{i} J \kappa}{2 \pi a \omega } \times
\begin{cases}
\frac{1}{1 - (\omega_{sp}/ \omega)^2 } ,&   n= \pm 1 \,\text{mod}\, J,  
\\
1 ,&  \text{otherwise}, 
\end{cases}
\\
\omega_{sp}^2 &= H_J \frac{\kappa}{m} , \quad H_J = 
\begin{cases}
J & J=1,2,
\\
J/2 & J \ge 3 . 
\end{cases}  
\end{aligned}
\end{equation}

The total scattering cross section (TSCS or $\sigma_{tot}$), which  is a measure of the power scattered in all directions due to an incident wave, will be used throughout this paper to compare various shell models. 
In the far-field $r\gg a$ the scattered pressure is $p_s = \sqrt{\frac{a}{2r}} \text{e}^{ikr} g(\theta) + O\Big( ( {kr})^{-3/2}\Big)$, where $g(\theta)$ is the form function, and hence 
\begin{equation} \label{sigma}
\sigma_{tot} = \frac{r}{a} \int\limits_{0}^{2\pi} p_s p_s^* d\theta  = \frac{1}{2} \int\limits_{0}^{2\pi} | g(\theta) |^2 d\theta  .
\end{equation}  
%= \frac{4}{ka} \sum\limits_{n=-\infty}^{\infty} | B_n |^2  

\section{Effective  properties of the shell-springs-mass system} \label{sec3}

For a given thin shell of thickness $h$, the internal springs increase the system stiffness and the added mass increases the density. 
This suggests the possibility of tuning the effective acoustic parameters of a stiffened shell through the values of stiffness $\kappa$ and mass $m$. 
The effective acoustic properties as functions of $\kappa$ and $m$ are determined next.

\subsection{Effective density}  \label{density}
 
%The total mass per unit axial length of the shell and internal rod is $m_s + m$. 
The effective density is the ratio of total mass to volume, $\rho_{eff} = (m_s + m)/(\pi a^2)$, where $m_s = \rho_s \pi (a^2 - (a-h)^2) $ is the mass per unit length of the shell. 
Hence, since $h\ll a$ by assumption, 
\begin{subequations} \label{1+1}
\begin{align}
\rho_{eff} &= \rho_m + \rho_s \Big( 2\frac ha - \big(\frac{h}{a}\big)^2 \Big)  \\
&\approx \rho_m + 2\frac{h}{a} \rho_s  , \ \ \text{where} \ \ \rho_m = \frac{m}{\pi a^2} .
\end{align}
\end{subequations}

\subsection{Effective bulk modulus} 

\begin{figure} [h!]
\begin{center}
\includegraphics[width=3.0in]{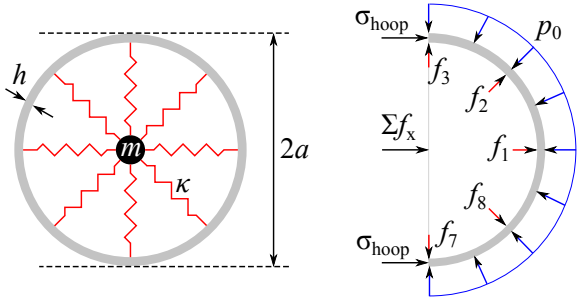}
\caption{External and internal forces acting on the shell with $J=8$ springs of stiffness $\kappa$ supporting a central mass $m$.} \label{sumF}
\end{center}
\end{figure}

Applying a hydrostatic pressure $p_0$ on the outside of the infinitely long cylindrical shell results in a decrease of the radius $a \to  a - \delta $. The quasi-static effective bulk modulus is defined as $K_{eff} = -{p_0}/({\Delta V/V})$, where the volume change is $\frac{\Delta V}{V}=\frac{\Delta A}{A}\approx -2\frac{\delta }{a} $ since the elastic deformation is plane strain. The radial and azimuthal strain in the shell are both approximately equal to $\varepsilon_{hoop}  = \frac{\delta }a$, so that the effective bulk modulus is  
\begin{equation} \label{2}
K_{eff} \approx \frac{p_0}{2\varepsilon_{hoop} }.
\end{equation}
The %plane strain nature of the problem also implies that the 
axial stress is $\sigma_a=\nu_s \sigma_{hoop}$, and consequently the hoop strain and stress are related by $\rho_s c_p^2 = \frac{E_s}{1-\nu_s^2}$ as
\begin{equation} \label{eps_hoop}
\varepsilon_{hoop} = \frac{1}{E_s} \left( \sigma_{hoop} - \nu_s \sigma_a \right) = \frac{\sigma_{hoop}}{\rho_s c_p^2} .
\end{equation}

An imaginary bisecting cut exposes the internal forces as shown in Fig.\ \ref{sumF}.  The %horizontal component of the 
static equilibrium condition is 
then 
 \begin{equation}
2h \,\sigma_{hoop} - 2a\,p_0  +F = 0 ,
\end{equation}
where  $F$ is  the horizontal resultant per unit length of the forces exerted by the springs on the half shell. 
At the same time, the spring forces are proportional to $\delta $, say 
\beq{3}
F = k_{eff} \delta   .
\eeq
Equations \er{2} - \er{3} imply that
\beq{4}
K_{eff}  = \frac{h}{2a}  \rho_s c_p^2  + \frac{ k_{eff}}4  ,
\eeq
where the effective stiffness $k_{eff}$ remains to be determined.  
Hence, referring to  Fig.\ \ref{sumF}, 
\beq{5}
F \equiv 
\sum f_x = \kappa \delta  \, \times 
\ca \sum\limits_{j=-J/4}^{J/4} \cos\theta_j  & \text{for even} \; J,
\\
 \sum\limits_{j=0}^{(J-1)/2} \sin\theta_j &  \text{for odd} \; J . 
\ac
\eeq
Performing the sums and using \er{3} gives 
\beq{52}
 k_{eff} = \kappa  \times 
\ca
 \cot\left(\frac{\pi}{J}\right)  & \text{for even} \; J,
\\
  \frac{1}{2} \cot\left(\frac{\pi}{2J}\right) &  \text{for odd} \; J .
\ac
\eeq

Consider  even $J$, in which case eqs.\ \er{4} and \er{52} imply that the effective bulk modulus is ($K_{sh}$ as in eq.\ (2.56) on page 38 of Ref.~\cite{Junger86})
\begin{equation} \label{Ke}
K_{eff} =  \frac{h}{a} K_{sh} + K_{sp} , 
\end{equation}
where $K_{sh} = \frac{\rho_s c_p^2}{2}$ and $K_{sp}=\frac{\kappa }4 \cot\big(\frac{\pi}J\big)$. The effective bulk modulus is greater than that of the bare shell and the increase, $K_{sp}$, is proportional to the spring stiffness. 
Substituting equations \eqref{1+1} and \eqref{Ke} into \eqref{Zsp}, the resonant frequency of the oscillator can be expressed in terms of the density and bulk modulus of the effective medium as
\begin{equation} \label{omega_sp}
\omega_{sp}^2 =   H_J \tan\left(\frac \pi J\right)\frac{4}{\pi a^2} \bigg( \frac{K_{eff}-
\frac{h}{a} K_{sh} }{\rho_{eff} - 2 \frac{h}{a} \rho_s } \bigg).
\end{equation}
%= H_J \tan \left(\frac \pi J\right) \frac{4}{\pi a^2} \frac{K_{sp}}{\rho_m}
%As $J$ becomes large $\tan(\pi/J) \rightarrow \pi/J$. 
In the case of odd $J$, $\tan(\pi/J)$ should be replaced by $2\tan(\pi/(2J))$ which has the same limit for large $J$, as expected. From here on the analysis will be carried out for $J$ even.

\section{Impedance and index matching} \label{sec4}
 
\subsection{Matched effective properties of water at low frequencies}

For a thin elastic shell to have the effective acoustic properties of water at low frequencies the internal oscillator must be tuned by selecting appropriate values of the added density $\rho_m$ and added bulk modulus $K_{sp}$ (see eqs.\ \er{1+1} and \er{Ke}). Setting the effective properties to water, $\rho_{eff}=\rho$ and $K_{eff}=K$, in eqs.\ \er{1+1} and \er{Ke} yields
\begin{equation} \label{cond}
\rho_m = \rho - 2 \frac{h}{a} \rho_s  \quad  \text{and} \quad  K_{sp} = K- \frac{h}{a} K_{sh} .
\end{equation}
The requirements that $\rho_m $ and $K_{sp} $ are non-negative impose an upper bound on the shell thickness: 
\begin{equation}
\frac{h}{a} < \frac{\rho}{2\rho_{s}}  \, \text{min}\Big(1,\ 
\Big( \frac{2c}{c_p}\Big)^2 \Big) , \qquad \frac{h}{a}\ll 1 .
%\times \begin{cases}
%1  &  c_p < 2c,   \frac{(2c)^2}{c_p^2}  &  c_p > 2c  . \end{cases}
\end{equation}

Note that if $c_p = 2c=(E_s/\rho_s (1-\nu_s^2) )^{1/2}$, then the empty shell has the same density and bulk modulus as water in the quasi-static limit. For typical engineering metals, $c_p > 2c$, therefore the realistic bound on shell thickness is a consequence of $K_{sp} \ge 0$, eq.\ \eqref{cond}$_2$. Table \ref{tab_mat} tabulates the upper bound on shell thickness for several materials. For some materials, the shell has to be thick in order to match the density of water,  so   thin-shell approximations do not apply. In that case equation \eqref{1+1} for density must be used exactly yielding an upper bound on thickness of $h/a=1-\sqrt{1-\rho/\rho_s }$. Similarly, tuning the bulk modulus to water requires the use of FEM for thicker shells.  

\begin {table}
\begin{center}
\begin{tabular}{| l | c | c | c | c || c | c |}  
\hline
Material & $\rho_s$ & $E_s$ & $\nu_s$ & $c_p$ & $\frac{h}{a}(\rho)$ & $\frac{h}{a}(K)$ \\ \hline 
\text{Al Oxide} & 3920 & 370 & 0.22 & 9959 & 0.137 & 0.012 \\   \hline
\text{Molybdenum} & 10300 & 276 & 0.32 & 5463 & 0.050 & 0.015 \\   \hline
\text{Al 3003-H18} & 2730 & 69 & 0.33 & 5326 & 0.204 & 0.058 \\ \hline
\text{Stl AISI 4340} & 7850 & 205 & 0.28 & 5323 & 0.067 &  0.020 \\   \hline
\text{Ti beta-31S} & 4940 & 105 & 0.33 & 4884 & 0.107 & 0.038 \\ \hline
\text{Copper} & 8700 & 110 & 0.35 & 3796 & 0.059 & 0.036 \\  \hline
\text{Concrete} & 2300 & 25 & 0.33 & 3493 & 0.248 & (0.143) \\   \hline
\text{Brick} & 2000 & 17 & 0.3 & 3056 & 0.293 & (0.199) \\   \hline
\text{Platinum} & 21450 & 147 & 0.39 & 2842 & 0.024 & 0.026 \\   \hline
\text{Silver} & 10500 & 72.4 & 0.37 & 2827 & 0.049 & 0.054 \\   \hline
\text{Acrylic} & 1190 & 3.2 & 0.35 & 1751 & 0.600 & (0.615)  \\   \hline
\text{ABS} & 1040 & 2.3 & 0.35 & 1588 & 0.804 & (0.761)  \\   \hline
\text{Lead} & 11340 & 13.87 & 0.42 & 1215 & 0.045 & (0.205) \\   \hline
\end{tabular}
\caption {The upper bound on shell thickness based on matched density $h/a (\rho_{eff}=\rho)$ and matched bulk modulus $h/a (K_{eff}=K)$ to water using several materials. Units of density are kg/m$^3$, elastic modulus GPa, speed m/s. The effective density is matched to water $\rho$ when $h/a=1-\sqrt{1-\rho/\rho_s }$. The quantities in parenthesis are outside the realm of thin shell theory and were obtained using FEM.}\label{tab_mat}
\end{center}
\end {table}

\begin{figure}
\begin{center}
\includegraphics[width=3.3in]{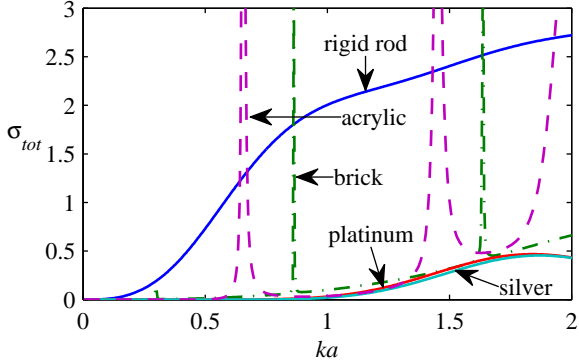}
\caption{Total scattering cross section vs.\ $ka$ for brick, acrylic, silver and platinum shells tuned to water with  $a=1$ cm and  $h/a=0.220, 0.615, 0.052, 0.0252$, respectively. } \label{brick}
\end{center}
\end{figure}

It is intriguing to see that for some materials such as platinum and silver the thickness bounds are nearly the same implying that at that thickness the shell has the stiffness and density of water. To investigate the effectiveness, simulations were done for plane wave incidence on shells of radius $a$=1 cm. The total scattering cross sections in Fig.\ \ref{brick} are negligible at low frequencies for brick, acrylic, silver and platinum shells of thickness $h/a=0.220, 0.615, 0.052, 0.0252$, respectively.  The total scattering cross section of a rigid rod of radius 1 cm is shown for comparison. There are several flexural resonances for the thick brick shell at $ka=0.297(n=2), 0.863(n=3), 1.638(n=4)$ and for the thick acrylic shell at $ka=0.658(n=2), 1.452(n=3)$, but none for the thin silver and platinum shells. 

Although thin silver and platinum shells exhibit transparency in water up to $ka=1$ they are not practical. For shells made of common engineering materials (i.e aluminum), stiffness and mass need to be added to obtain the density and bulk modulus of water. A procedure for  tuning thin elastic shells is described next.

\subsection{Tuning an aluminum shell to water} \label{asiw}

\subsubsection{Internal resonance}

\begin{figure} [H]
\begin{center}
\includegraphics[width=3.3in]{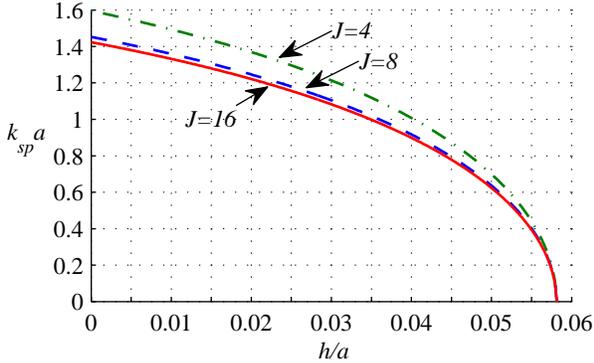}
\caption{Non-dimensional resonant wave number $k_{sp}a = \omega_{sp}a/c$ of the internal springs-mass system as a function of shell thickness (see eq.\ \eqref{omega_sp}). The aluminum shell is tuned to water.} \label{res_ha_alum}
\end{center}
\end{figure}

Consider a thin shell made of aluminum 3003-H18 ($\rho_s = 2730\,\text{kg/m}^3$, $c_p = 5326\, \text{m/s}$). Matching the effective properties of the shell-springs-mass system to water as $K_{eff}=K$ and $\rho_{eff}=\rho$, yields  the 
resonant frequency $\omega_{sp}$ of eq.\ \eqref{omega_sp}
A necessary condition for low frequency matching of the effective properties is that the first internal resonance lies above the low-frequency range, here considered as roughly $0< ka < 0.5$.  
Figure \ref{res_ha_alum} plots the non-dimensional resonant wave number $k_{sp} a$ 
where  $k_{sp}=\omega_{sp}/c$. 
% for $J=4, 8, 16$ springs. 
As the shell thickness decreases, the resonant frequency of the internal oscillator increases at a diminishing rate. Equivalently,  as the shell becomes thinner the added stiffness must increase faster than the mass. Also note that the resonant frequency drops as the number of springs, $J$, increases. This is due to the factor $H_J \tan(\pi/J)$ in eq.\ \eqref{omega_sp}.

From Fig.\ \ref{res_ha_alum} as well as Table \ref{tab_mat} we see that the upper bound on shell thickness is $h/a=0.058$. Since aluminum is relatively light there is a substantial mass deficiency $\rho_m=0.683\rho$. In order to tune the shell to water ($\rho = 1000\,\text{kg/m}^3$, $c = 1500\, \text{m/s}$), a central mass is added. However, the mass has to be supported by springs so the shell must be thinner than the upper bound, namely $h/a=0.03$, to accommodate the additional stiffness.

\subsubsection{Flexural resonances}
\begin{figure}
\centering
\subfloat[]{ \includegraphics[width=3.2in]{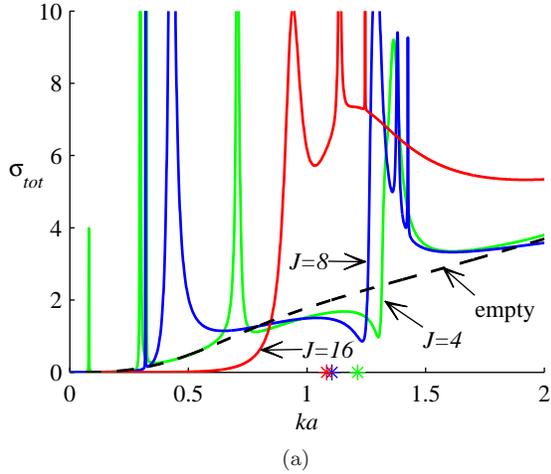}    \label{TSCSfull}  }
\\
\subfloat[]{ \includegraphics[width=3.2in]{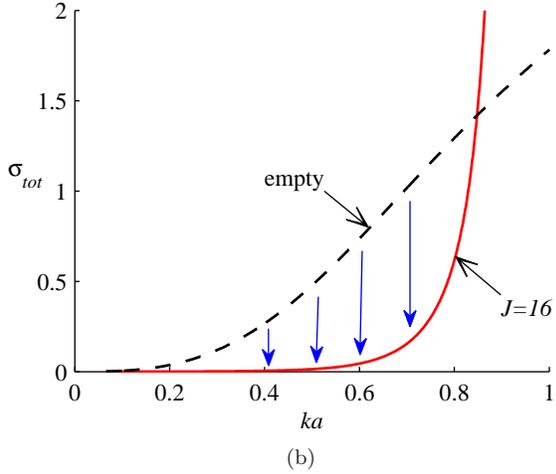}   \label{TSCSsmall}  }
\caption{Plot (a) shows the total scattering cross-section for an aluminium shell of thickness $h/a=0.03$ with $J=4,8,16$ springs supporting a central mass. The dashed line is the TSCS of the empty shell. Plot (b) is a close up of plot (a) showing the achieved decrease in the scattering cross-section from an empty shell to a tuned shell with $J=16$ springs.} \label{TSCS}
\end{figure}

The scattering response of the tuned shell is analyzed for plane wave incidence. % along the x-axis.
 Figure \ref{TSCSfull} shows the total scattering cross section (TSCS) of eq.\ \eqref{sigma} as a function of $ka$ for the three cases in Fig.\ \ref{res_ha_alum} at the thickness ratio $h/a = 0.03$. The star on the horizontal axis indicates the resonant frequency of the springs-mass system, which per previous discussion decreases with $J$. For $J=4$ and $J=8$ springs the tuning is only effective at extremely low frequencies, because of the presence of several flexural resonances. However with $J=16$ springs the TSCS is close to zero at frequencies up to $ka=0.8$, where the magnified view is shown in Fig.\ \ref{TSCSsmall}. By further increasing the number of springs, the transparent region increases only slightly, because it is bounded by the resonant frequency of the oscillator which for large $J$ is at about $k_{sp}a=\omega_{sp}a/c = 0.95$ (see eq.\ \eqref{omega_sp}).

The shell thickness of $h/a=0.03$ is the optimal shell thickness, because it maximizes the range of frequencies of low TSCS. At the optimal thickness the lowest resonance of the combined acoustic and shell impedances $\big(\sum\limits_{p = - \infty}^\infty 1/(Z^{sh}_{n+pJ}+Z_{n+pJ}) \big)^{-1}$ coincides with the resonant frequency of the oscillator $Z_n^{sp}$. The result is that there is a large region free from flexural resonances but still close enough to the oscillator resonance for it to be effective. 

\begin{figure} [H]
\centering

\subfloat[J=4, n=3]{   \includegraphics[width=1.5in] {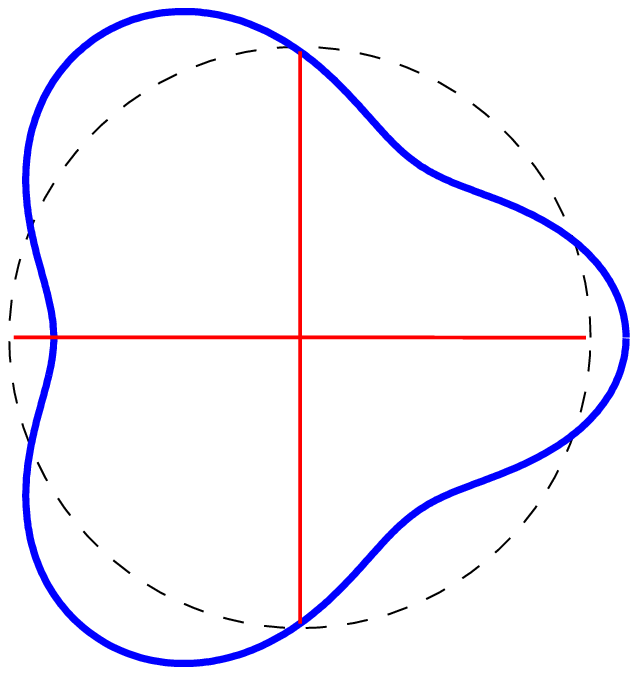}     \label{J4_n3}  } 
\subfloat[J=4, n=5]{   \includegraphics[width=1.5in] {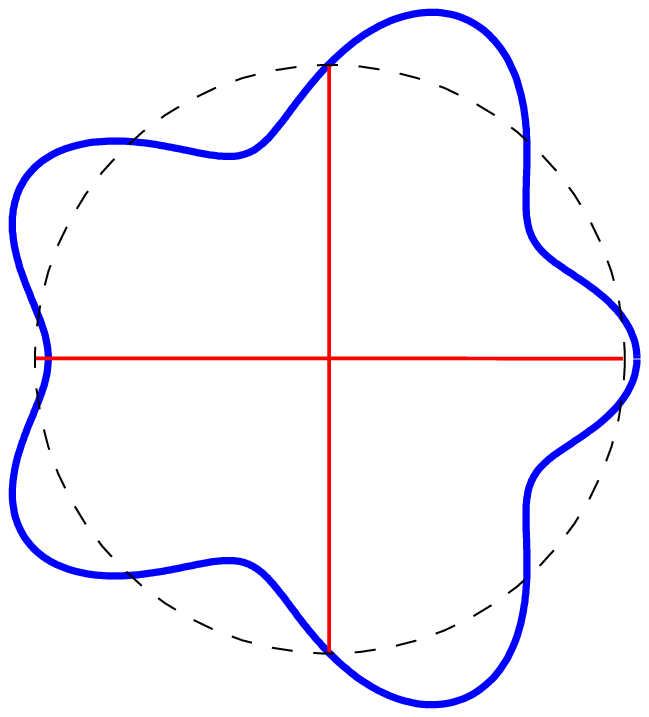}     \label{J4_n5}  } 
\\
\subfloat[J=4, n=6]{   \includegraphics[width=1.5in] {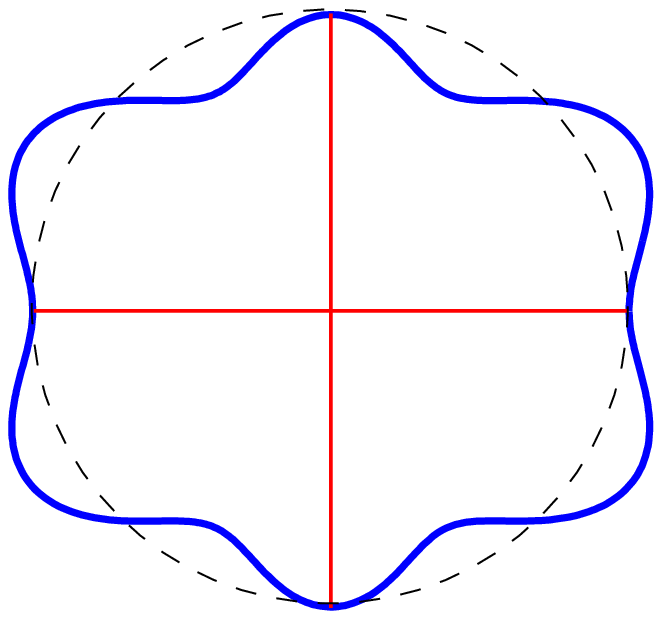}     \label{J4_n6}  } 
\subfloat[J=4, n=7]{   \includegraphics[width=1.5in] {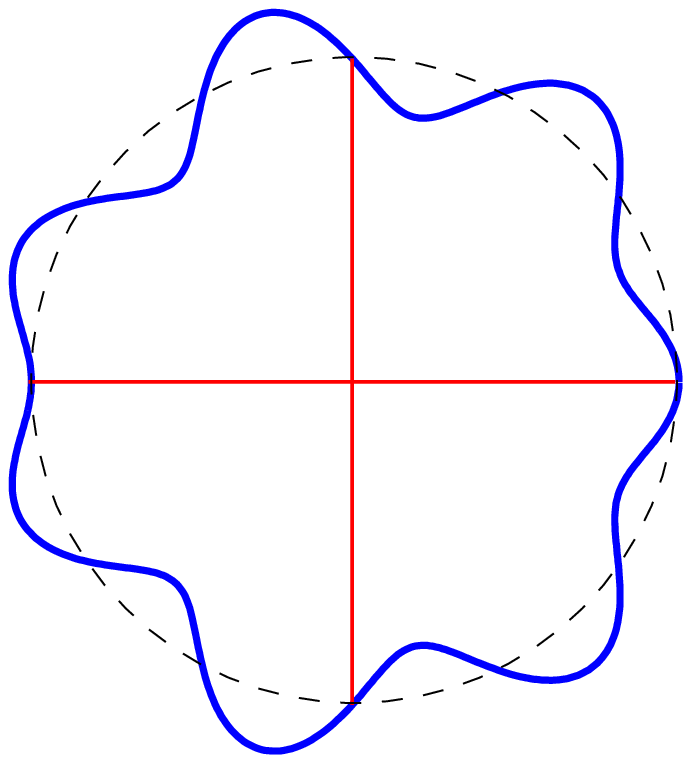}     \label{J4_n7}  }
\\
\subfloat[J=8, n=5]{   \includegraphics[width=1.5in] {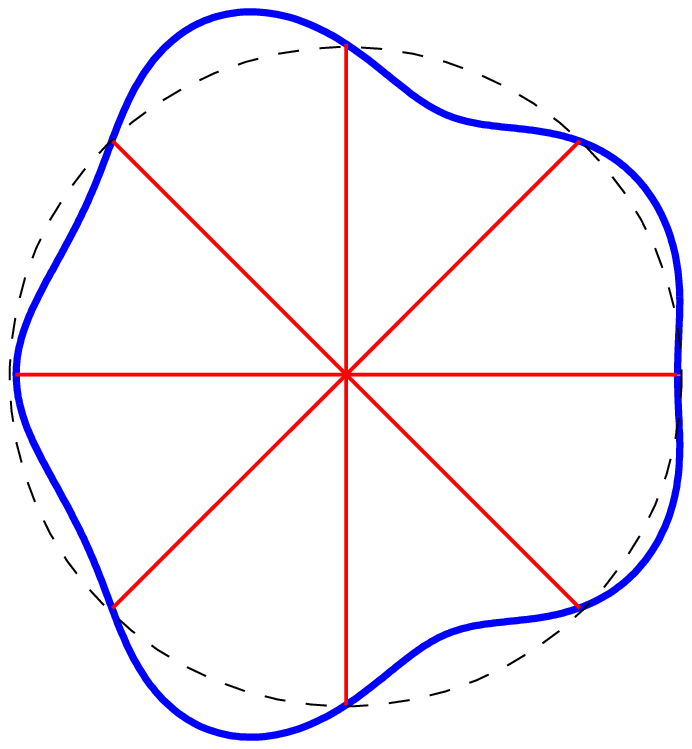}     \label{J8_n5}  } 
\subfloat[J=8, n=6]{   \includegraphics[width=1.5in] {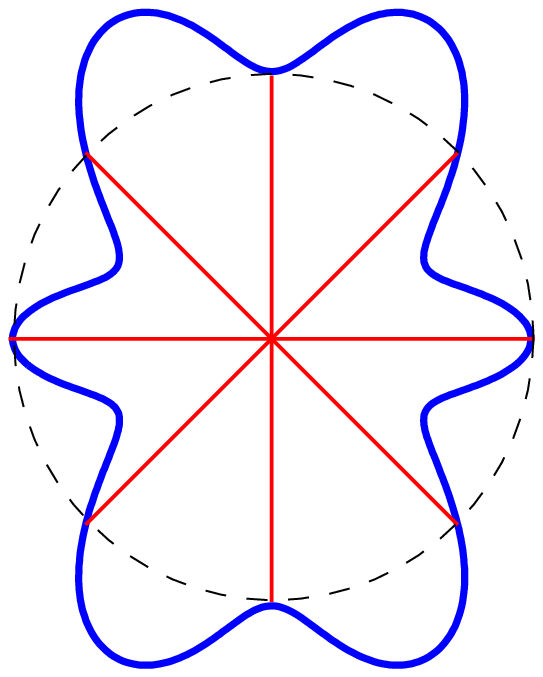}     \label{J8_n6}  } 
\\
\subfloat[J=8, n=7]{   \includegraphics[width=1.5in] {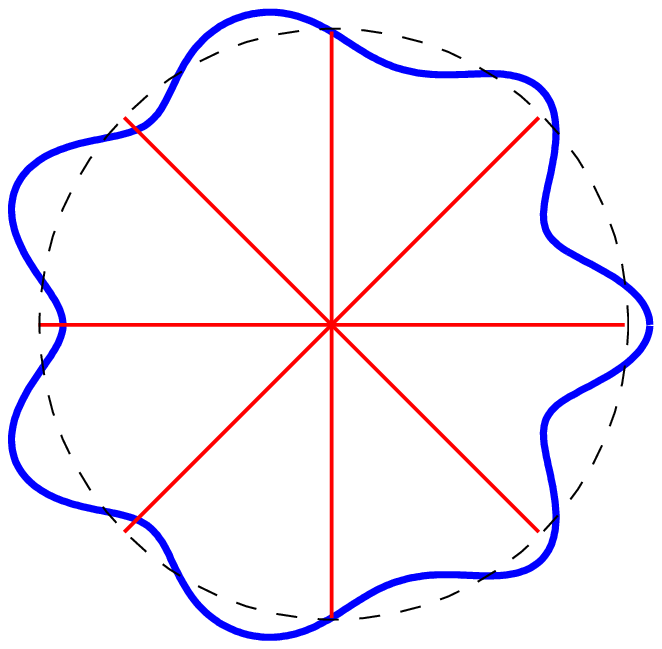}     \label{J8_n7}  }
\subfloat[J=16, n=15]{ \includegraphics[width=1.5in] {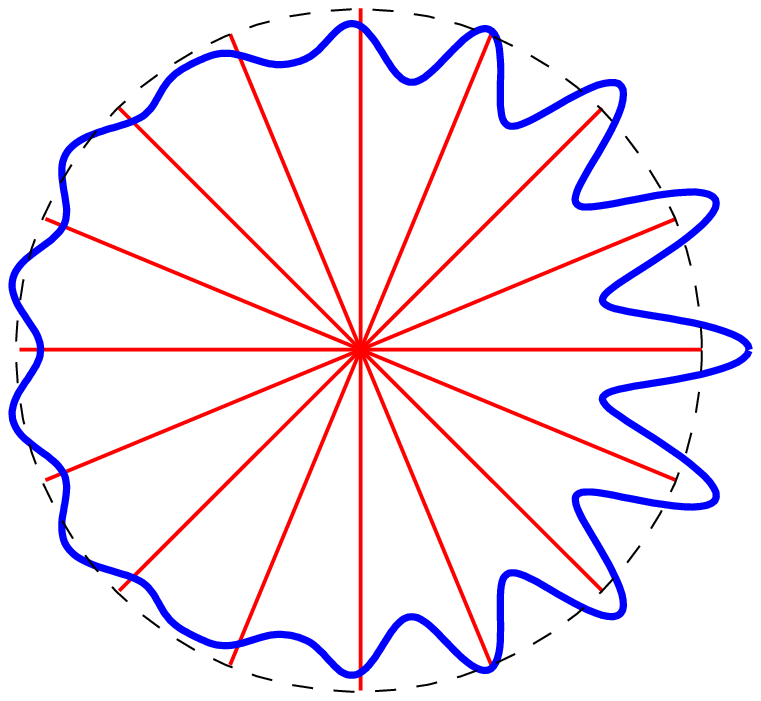}       \label{J16_n15}}
\caption{Radial displacement $w(\theta)$ for the $J=4$ case at resonant frequencies: $ka=0.081,0.303,0.338,0.74$ in (a), (b), (c) and (d), respectively. Plots (e), (f) and (g) show the radial displacement $w(\theta)$ for the case with $J=8$ springs at resonant frequencies: $ka=0.273,0.321,0.434$, respectively. Plot (h) is the radial displacement for $J=16$ springs at the resonant frequency of $ka=0.945$. The radial lines depict the internal springs. The thickness of the aluminum shell is $h/a=0.03$. Displacement has been arbitrarily scaled for clear depiction of the mode shape.}
\label{fig5}  
\end{figure}

The reason for the decrease in the number of flexural resonances with increasing $J$ can be understood by considering the radial displacement of the shell $w(\theta)$ at each resonance. These are plotted in Fig.\ \ref{fig5}, where indeed each resonance corresponds to a certain flexural mode. The red radial lines indicate the positions of the springs. From these we can conclude that as the number of springs $J$ increases more flexural modes are constrained by the springs. The modes that do appear are either modes where the spring attachments coincide with the anti-nodes of the radial displacement or if the mode is odd, the displacement is anti-symmetric.

Note that although we can attribute each resonance peak to a flexural mode, the position of the peak is difficult to predict in the low frequency range, see Ref.~\cite{Titovich13} for further details. %Titovich and Norris did obtain an effective resonant frequency for moderate frequencies

%\begin{figure}[H]    
%\centering
% \hspace{-.5in}
%\subfloat[J=4, n=3]{   \includegraphics[width=1.5in] {figures/resonances/J_4_ka_0081_n3.eps}     \label{J4_n3}  } \hspace{-.55in}
%\subfloat[J=4, n=5]{   \includegraphics[width=1.5in] {figures/resonances/J_4_ka_0303_n5.eps}     \label{J4_n5}  } 
%\\
%\hspace{-.55in}
%\subfloat[J=4, n=6]{   \includegraphics[width=1.5in] {figures/resonances/J_4_ka_0338_n6.eps}     \label{J4_n6}  } \hspace{-.55in}
%\subfloat[J=4, n=7]{   \includegraphics[width=1.5in] {figures/resonances/J_4_ka_0740_n7.eps}     \label{J4_n7}  }
%\\
%\hspace{-.5in}
%\subfloat[J=8, n=5]{   \includegraphics[width=1.5in] {figures/resonances/J_8_ka_0273_n5.eps}     \label{J8_n5}  } \hspace{-.55in}
%\subfloat[J=8, n=6]{   \includegraphics[width=1.5in] {figures/resonances/J_8_ka_0321_n6.eps}     \label{J8_n6}  } 
%\\
%\hspace{-.55in}
%\subfloat[J=8, n=7]{   \includegraphics[width=1.5in] {figures/resonances/J_8_ka_0434_n7.eps}     \label{J8_n7}  } \hspace{-.55in}
%\subfloat[J=16, n=15]{ \includegraphics[width=1.5in] {figures/resonances/J_16_ka_0945.eps}     \label{J16_n15}  }
%\caption{First row: Radial displacement $w(\theta)$ for the $J=4$ case at resonant frequencies: $ka=0.081,0.303,0.338,0.74$ in a, b, c and d, respectively. Second row: Radial displacement $w(\theta)$ for the $J=8$ and $J=16$ cases at resonant frequencies: $ka=0.273,0.321,0.434$ and $ka=0.945$ in e, f, g, h, respectively.  The red lines depict the internal springs. Shell thickness is $h/a=0.03$. Displacement has been arbitrarily scaled for clear depiction of the mode shape.}
%\label{fig5}  
%\end{figure}

\section{Internal oscillator designs} \label{sec5}

The analytical model demonstrated the theoretical possibility of tuning elastic shells. We next consider several physical oscillator designs selected for their effectiveness and ease of manufacturing. 

\subsection{A one-component internal oscillator} \label{elast1}
\begin{figure} [h]
\begin{center}
\includegraphics[width=1.7in]{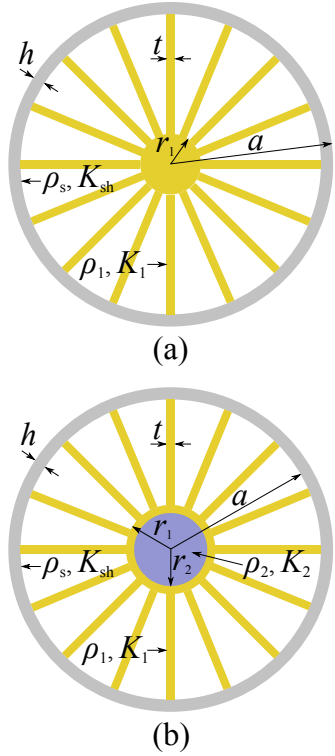}
\caption{Schematic of the tuned shell. Plot (a): a one-component internal mechanism consists of $J=16$ stiffeners (ribs) with thickness $t$ and a central mass (rod) of radius $r_1$. Plot (b) shows the same internal mechanism, but with an added internal rod of radius $r_2$.} \label{shell_insert}
\end{center}
\end{figure}

Consider a one-component internal mechanism consisting of $J=16$ elastic stiffeners (ribs) of thickness $t$ and a central mass (rod) of radius $r_1$ made of the same material as shown in Fig.\ \ref{shell_insert}(a).
The elastic modulus and density of the internal mechanism are $E_{1}$ and $\rho_{1}$, respectively. Assuming that the stiffeners only deform axially, the effective stiffness (per unit axial length) %is proportional to the thickness and length of the stiffener and 
is  $\kappa={E_{1} t}/(a-h-r_1)$.  This first order approximation for additional stiffness $\kappa$ will prove sufficient for low frequency tuning. The second parameter of interest is the mass of the internal mechanism,  $m_{1} = \rho_{1} \left( J t (a-h-r_1) + \pi r_1^2  \right)$.

The two variables which define the geometry of the internal mechanism, $t$ and $r_1$, determine the effective bulk modulus and density of the shell-oscillator system. From \eqref{Ke}, the contribution of the internal oscillator to the bulk modulus of the shell-stiffener-mass system is $K_{sp}=\frac 14 {\cot(\pi/J)E_{1} t}/(a-h-r_1)$.  Thus the effective properties of the combined system become (see eqs.\ \eqref{Ke} and \eqref{1+1})
\begin{subequations} \label{cond1}
\begin{align}
K_{eff} &= K_{sh} \frac{h}{a} + \frac{\cot(\pi/J)E_{1} t}{4(a-h-r_1)}  ,  \label{Kapp}
\\ 
\rho_{eff} &= \rho_{sh} \frac{h}{a}\Big(2- \frac{h}{a}  \Big) 
\\
&+ \rho_{1}\Big(\frac{J}{\pi}\frac{t}{a}\big(1-\frac{h}{a}-\frac{r_1}{a} \big) + \big(\frac{r_1}{a} \big)^2 \Big) ,  \label{rapp}
\end{align}
\end{subequations}
where the O($(h/a)^2$) term  in the shell volume is retained for improved accuracy. To get the effective properties, the density of the shell has to be increased by $\rho_m=\rho_{eff} - \rho_{sh} \big(2 \frac{h}{a} - (\frac{h}{a})^2 \big)$. For example, from Table \ref{tab_mat} the mass deficiency for an aluminum shell with matched compressibility in water is $\rho_m=0.683\rho$.

Next define the ratio of required additional stiffness in each stiffener to the elastic modulus of the internal material 
\begin{equation}\label{17}
\hat{K} = \frac{4 J}{\pi E_1} \big(K - K_{sh} \frac{h}{a} \big)\tan\big(\frac{\pi}J \big) .
\end{equation}
Solving the compressibility condition \eqref{Kapp} for $t$ yields 
\begin{equation} \label{t}
t = J^{-1}{\pi \hat{K}} (a-h-r_1) .
\end{equation}
Substituting   into the density condition in \eqref{rapp}     yields   a quadratic equation for $r_1/a$,
\begin{equation} \label{quad_rc}
  \Big( \frac{r_1}{a} \Big)^2 + \hat{K}\Big( 1-\frac{h}{a} - \frac{r_1}{a} \Big)^2 
- \frac{\rho_m}{\rho_1} = 0 .
\end{equation}
For $h/a\ll 1$, the roots are approximately ${r_1}/{a} =(\hat{K}+1)^{-1} \big(
\hat{K} \pm \sqrt{ (\rho_m / \rho_1)(\hat{K}+1)-\hat{K} } \big)$.

\subsection{A two-component internal oscillator} \label{elast3}

It can happen that  the density of the internal mechanism material is so low, as in some plastics, that it becomes difficult to match both the density and bulk modulus. In that case, a heavy central rod of radius $r_2$ can be added as shown in Fig.\ \ref{shell_insert}(b). This rod has practically no effect of the effective stiffness, but does offset the density. The effective bulk modulus for this system is the same as in equation \eqref{Kapp}, and the effective density changes as follows
\begin{multline} \label{cond_new} 
\rho_{eff} =\rho_{sh} \frac{h}{a}\Big(2- \frac{h}{a}  \Big) 
\\
+ \rho_{1}\Big(\frac{J}{\pi}\frac{t}{a}\big(1-\frac{h}{a}-\frac{r_1}{a} \big) + \big(\frac{r_1}{a} \big)^2 \Big) + (\rho_2 -\rho_1)\Big(\frac{r_2}{a}\Big)^2  ,
\end{multline}
where $\rho_2$ and $K_2$ are the density and bulk modulus of the central rod. Three parameters now  define the geometry of the internal mechanism: $t$, $r_1$ and $r_2$. Since there are two conditions,  \eqref{Kapp} and \eqref{cond_new}, the radius of the internal oscillator is determined after selecting the fraction of density added by the rod $f_{\rho} \le \rho_m/\rho_2$, which yields $r_2/a=\sqrt{f_{\rho}}$
Recalling eq.\ \eqref{t} for $t$, and using the definition of $\rho_m$, equation \eqref{cond_new}  can be  rearranged as 
\begin{equation} \label{quad_rc2}
 \Big( \frac{r_1}{a} \Big)^2 + \hat{K}\Big( 1-\frac{h}{a} - \frac{r_1}{a} \Big)^2 
+ \Big(\frac{\rho_2}{\rho_1} - 1\Big) \Big(\frac{r_2}{a}\Big)^2 - \frac{\rho_m }{\rho_1} = 0. 
\end{equation}
It is clear if $\rho_2=\rho_1$, equation \eqref{quad_rc2} gives the solution for a one-component oscillator, i.e.\ eq.\ \eqref{quad_rc}. For a very thin shell with $h/a\ll 1$, the roots are approximately 
${r_1}/{a} =(\hat{K}+1)^{-1} \big(
\hat{K} \pm \sqrt{ (\hat{\rho}/ \rho_1)(\hat{K}+1)-\hat{K} } \big)$,
 where $\hat{\rho} = \rho_m - f_{\rho}(\rho_2-\rho_1)$. 
The discriminant goes to zero if we select $f_{\rho}$ such that $\frac{\rho_m}{\rho_1}-f_{\rho}(\frac{\rho_2}{\rho_1}-1) = \frac{\hat{K}}{\hat{K}+1}$ giving the single solution $\frac{r_1}{a} = \frac{\hat{K}}{\hat{K}+1}$ which  corresponds to the largest possible central rod.

\subsection{Aluminum shell with an acrylic internal mechanism}
The shell is made of Aluminum 3003-H18   ($\rho_s=2730\,\text{kg/m}^3$, $E=69\,\text{GPa}, \nu = 0.33$) and the  oscillator is acrylic ($\rho_{1}=1190\,\text{kg/m}^3$, $E_{1}=3.2\,\text{GPa}, \nu = 0.35$). The longitudinal speed of sound,  $c_l = 
c_p (1-\nu)/\sqrt{ 1-2\nu}$, is $6120$ m/s for aluminum and $2078$ m/s for acrylic. As in the previous simulations, the shell has  outer radius of $a=1$ cm and thickness $h/a=0.03$. Using $J=16$ stiffeners the shell is tuned such that the effective properties mimic water.

We first solve equation \eqref{quad_rc} for $(r_1/a)$ and then apply equation \eqref{t} to get the thickness, giving two solutions for the oscillator's parameters $(r_1, t) = (8.21,0.40) $ and $(3.03, 1.80)$ (in mm). Note that the lower limit on the internal mass radius $r_1$ is geometrically constrained by the thickness of each stiffener. The intersection of stiffeners gives a lower bound of roughly $r_1>Jt/(2\pi)$. In the second solution the radius $r_1$ is below this bound. This implies that the density is not matched to water and consequently only the first solution is retained.

Although the present representation of the added stiffness is a good approximation it is not exact. The solutions  were optimized in COMSOL yielding the exact bulk modulus of water. The geometry of the oscillator was found to be
\begin{equation} \label{roots_opt}
(r_1, t) = (7.96,0.81)\,\text{mm}  
\end{equation}
corresponding to  $(\rho_{eff}, K_{eff}) = (1000.8\,\text{kg/m}^3,2.251\,\text{GPa})$. 

The radius $r_1$ is large compared to the shell radius $a$. It can be made smaller while still matching the effective properties to water by inserting a central steel rod ($\rho_s=7944\,\text{kg/m}^3$, $E=200\,\text{GPa}, \nu = 0.28$). Solving equation \eqref{quad_rc2} and using \eqref{t} yields $(r_1, r_2, t) = (5.60, 1.67, 1.10)$mm, which were optimized in COMSOL to give 
\begin{equation} \label{roots_rod_opt}
(r_1, r_2, t) = (5.60, 1.29, 1.43)\,\text{mm} .
\end{equation}

\begin{figure} [H]
\begin{center}
\includegraphics[width=3.3in]{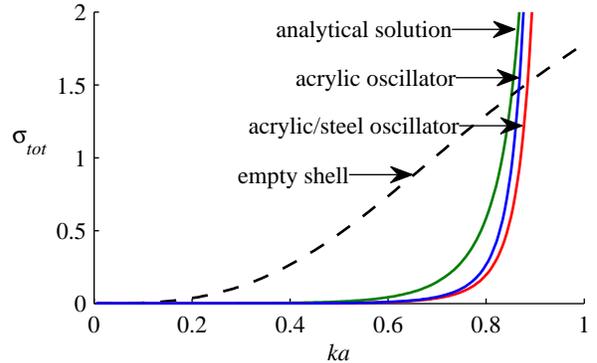}
\caption{Comparison of the TSCS for different oscillator designs for an aluminum shell of thickness $h/a=0.03$.} \label{P_ratio} 
\end{center}
\end{figure}

Figure \ref{P_ratio} shows the TSCS for the two oscillator designs in eqs.\ \eqref{roots_opt} and \eqref{roots_rod_opt} as well as the analytical springs-mass solution of Fig.\ \ref{TSCSsmall} and that for the empty shell. The FEM simulations were carried out in COMSOL.
The presence of the oscillator significantly decreases the scattered power at low frequencies. The TSCS is effectively zero at frequencies below $ka=0.6$ making the shell transparent in water. The oscillator with the central rod gives the broadest region of negligible scattering. The accuracy of  the analytical springs-mass solution~\cite{Titovich13} vs.\ the full  FEM simulations, evident in Fig.\ \ref{P_ratio}, is quite remarkable.

\section{A cylindrical to plane wave lens constructed from an array of tuned shells} \label{sec6}
In this section we consider a relatively small (7x7) array of tuned shells demonstrating  wave steering capabilities.

\subsection{Unit cell of a fluid saturated array of shells} 
\begin{figure}[H]    
\centering
\includegraphics[width=1.9in]{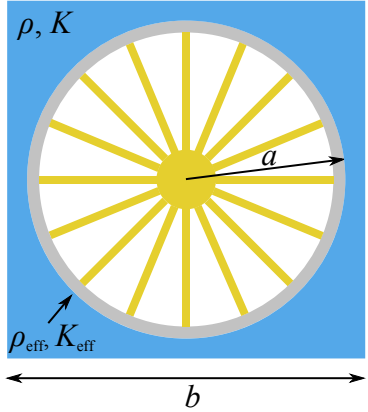}  
\caption{A square unit cell of a fluid saturated array of shells.} \label{unit} 
\end{figure}

The unit cell of the square array, shown in Fig.\ \ref{unit}, consists of a central shell-springs-mass system surrounded by a square region of water.
The shell volume fraction in the unit cell is $f=\pi a^2/b^2$, where $b$ is the cylinder spacing as well as the side length of the unit cell. The equivalent density and bulk modulus, $\rho_{eq},K_{eq}$, of the unit cell depend on the surrounding fluid as
%\begin{subequations} \label{Kcell}
%\begin{align}
%\frac{\rho_{eq}}{\rho}  = 1 + f \Big(\frac{\rho_{eff}}{\rho} -1 \Big)  \quad &\Rightarrow \quad  \frac{\rho_{eff}}{\rho}  = 1 + \frac{1}{f} \Big(\frac{\rho_{eq}}{\rho} -1 \Big) ,
%\\
%\frac{K_{eq}}{K} = \frac{1}{1+f(\frac{K}{K_{eff}}-1)}  \quad &\Rightarrow  \quad  \frac{K_{eff}}{K} = \frac{1}{1+\frac{1}{f}(\frac{K}{K_{eq}}-1)}  .
%\end{align}
%\end{subequations}
\begin{subequations} \label{Kcell}
\begin{align}
\frac{\rho_{eq}}{\rho}  = 1 + f \Big(\frac{\rho_{eff}}{\rho} -1 \Big) ,
\\
\frac{K_{eq}}{K} = \frac{1}{1+f(\frac{K}{K_{eff}}-1)} .
\end{align}
\end{subequations}
The equivalent density and bulk modulus of the unit cell are significantly affected by the surrounding fluid. For shells of radius $a=1$cm with a relatively tight packing of $b=2.2a$ yields a filling fraction of $f=0.65$. In this case, in order to have the effective quasi-static bulk modulus of the unit cell $K_{eq}=2K$, the effective bulk modulus of the shell-springs-mass system must be $K_{eff} = 4.33K$.

The effective impedance of each shell relative to water (acoustic impedance $Z = \sqrt{\rho K}$)  is determined by 
\begin{equation}
\frac{\rho_{eff} K_{eff}}{\rho K} = \frac{\rho_{eq}K_{eq}}{\rho K} \Big( \frac{1-(1-f)\rho / \rho_{eq}}{1-(1-f)K_{eq} / K} \Big) .
\end{equation}

\subsection{Bulk modulus distribution via conformal map} 
The wave equation for an acoustic medium is invariant under coordinate transformations. Moreover, if the transformation 
$\gamma = x+iy\rightarrow s = x'+iy'$ is conformal, $s = s(\gamma)$, then the mapped density $\rho'$  and bulk modulus  $K'$  in the transformed coordinates are \cite{Norris12a} 
\begin{equation} \label{map}
\rho' =  \rho, \qquad K' =  K | ds /d\gamma| .
\end{equation}  

We consider the conformal transformation  of a unit  $\gamma$ circle to a unit $s$ square. The   circle is first mapped to the upper half plane through a bilinear transformation; the subsequent polygon mapping  takes the upper half plane to the unit square in  $s$. The resulting unit square to  unit circle inverse mapping  is 
\begin{subequations} \label{1-15b}
\begin{align}
\gamma &= \frac{1-\chi}{1+\chi} e^{-i\pi/4},
\\
\chi &= i\,\text{cn}^2\left(\frac 12 
\bold{K} \left(\frac 1{\sqrt{2}}\right)( s +1+i) \right) ,
\end{align}
\end{subequations}
where $\bold{K}()$ is the complete elliptic integral of the first kind and $\text{cn}(u)$ is the Jacobi elliptic function. The bulk modulus distribution in the transformed space is 
\begin{equation}\label{1-15}
K' %= K s ' (\gamma)
=\frac{2 K}{ \bold{K}(\frac 1{\sqrt{2}})\sqrt{\gamma^4+1 }} .
\end{equation}

\begin{figure*}[h!] 
\centering
\subfloat[]{   \includegraphics[width=3.0in, height=3in]{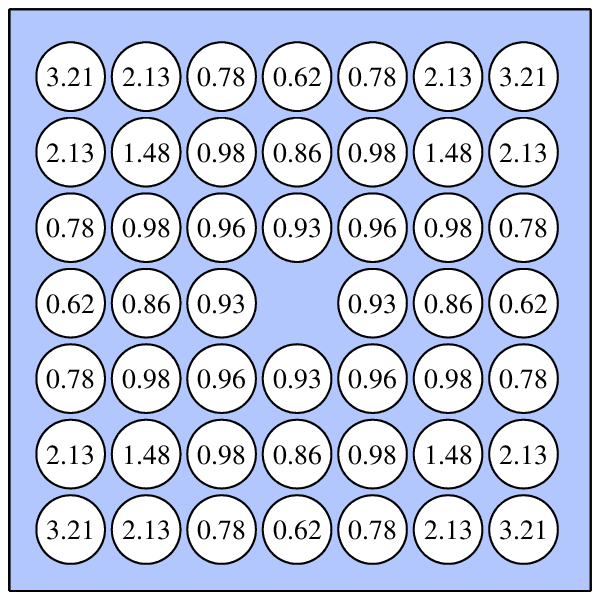}     \label{distrib}  }
\subfloat[]{   \includegraphics[width=3.0in, height=3in]{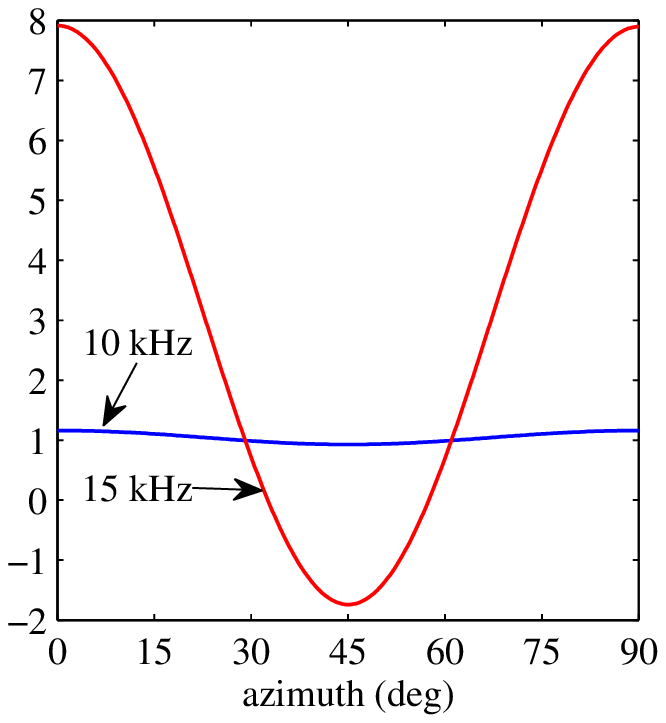} \label{pres_line}  } 
\\
\subfloat[]{   \includegraphics[width=3.0in, height=3in]{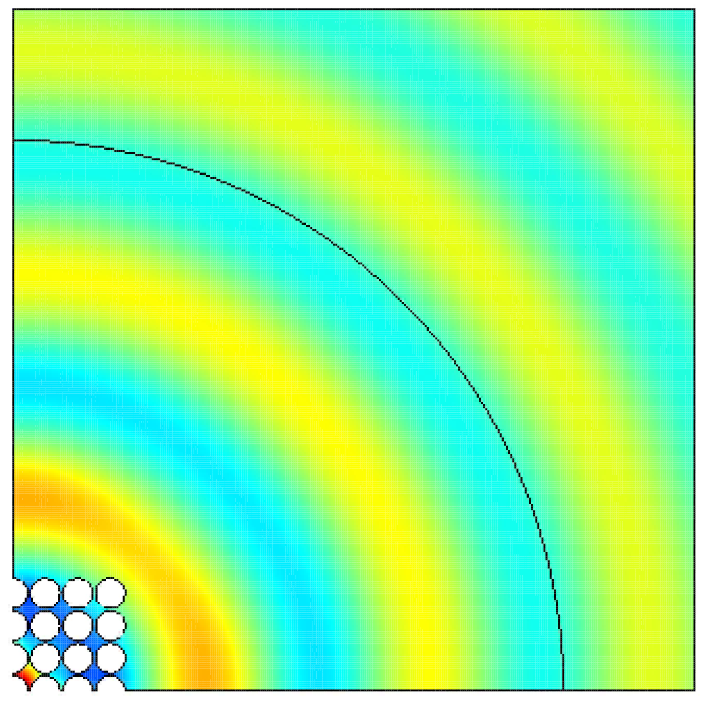}  \label{10kHz}  }
\subfloat[]{   \includegraphics[width=3.0in, height=3in]{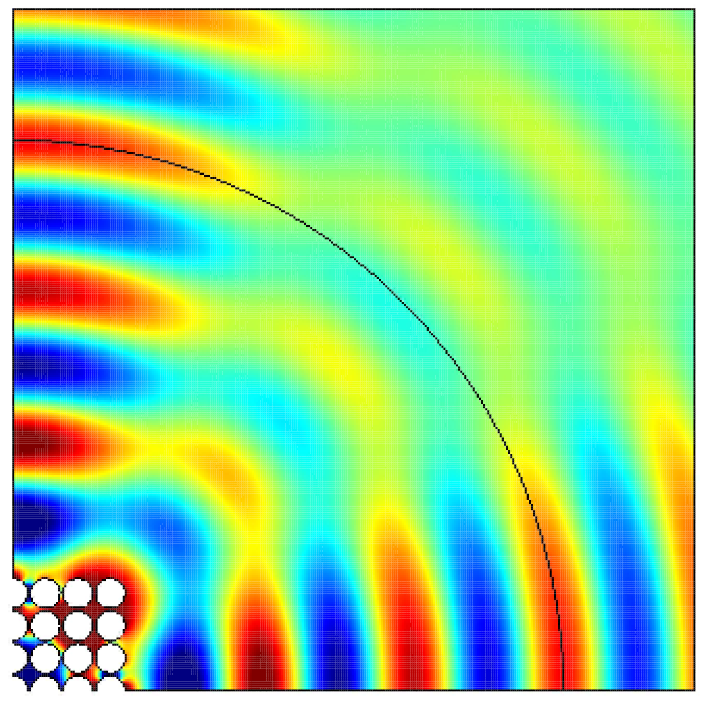}  \label{15kHz}  } 
\caption{Cylindrical to plane wave lens. Plot (a) shows the bulk modulus distribution in the 7x7 array of tuned shell. Plots (c) and (d) show the pressure field around the lens at 10kHz and 15kHz, respectively. Plot (b) is the pressure normalized by the monopole source pressure without the lens along the quarter circular arcs in (c) and (d).}
\label{fig9}  
\end{figure*}

The distribution \eqref{1-15} is used to design a cylindrical-to-plane wave lens. The proposed array contains 7x7 unit cells of size $b=2.2a$ with cylinders of radius $a=1$cm giving a filling fraction $f=0.65$ and the side length of the lens $L=15.4$cm. The complex variable defining the square is $\{s(x+iy) | x,y:{-L/2,L/2} \}$. Substituting $s$ into \eqref{1-15b} and then the obtained $\gamma$ into \eqref{1-15} gives the continuous function of the bulk modulus distribution. This function is then discretized by averaging it over each unit cell. Using this required equivalent stiffness of each unit cell $K_{eq}$ the properties of the shell-springs-mass system are obtained from eq. \eqref{Kcell} as $\frac{K_{eff}}{K} = (1+1.54(\frac{K}{K_{eq}}-1))^{-1}$. Figure \ref{distrib} shows the bulk modulus of each shell-spring-mass system normalized to water. The effective density of each system is tuned to water, see \eqref{map}.

Each shell-spring-mass system is designed by the method outlined in Section \ref{sec5}. The thickness of the aluminum shells has to vary form $0.03$ to $0.12$ to achieve this inhomogeneity of bulk modulus from $\frac{K_{eff}}{K}=0.93 \; \text{to} \; 3.21$. Appropriate geometry of acrylic internal oscillator with $J=16$ stiffeners tunes the shell to the required acoustic properties. The slow shells with $\frac{K_{eff}}{K}$ ranging from 0.62 to 0.86 are made of acrylic with $\frac{h}{a} = 0.3$ and tuned with an acrylic oscillator. The central shell is removed to give room for a monopole source. 

The total pressure field was obtained by simulating the lens made of elastic shells in COMSOL. A symmetric quarter of the pressure field at monopole source frequencies of 10 kHz ($ka=0.42, \lambda/a=15$) and 15 kHz ($ka=0.63, \lambda/a=10$) is shown at same scales in Figures \ref{10kHz} and \ref{15kHz}, respectively. Also, Figure \ref{pres_line} shows the pressure ratio along the quarter circular arcs in Figures \ref{10kHz} and \ref{15kHz} between the field with the lens (shown) and source only (not shown).

At low frequencies the wavelength is much larger than the shell size $\lambda/a=15$ and the lens is essentially transparent. However, at 15 kHz when $\lambda/a=10$ each tuned shells behaves as an effective acoustic medium steering the wave from the corner to the faces. The wave travels across only 3 rows of shells and the maximum amplitude is magnified by a factor of 7 as seen in Figure \ref{pres_line}. The increase in the pressure amplitude from the faces and its decrease from the corner demonstrates wave steering. 

A larger array of shells will increase the effectiveness of the lens. In the design of each shell, it is important to understand that this is a model with three parameters: shell thickness, oscillator stiffness and mass. The effectiveness of the internal oscillator changes with shell thickness. The following procedure will guarantee a successful design of an acoustic lens:
\begin{itemize}
  \item Select the thickness of each shell to optimize the range of frequencies for it to behave as an effective medium. 
  \item Design each oscillator so as to suppress all low frequency flexural waves of the shell and maximize its natural frequency.
\end{itemize}

\section{Conclusions}  \label{concl}
Elastic shells can be tuned to yield quasi-static effective acoustic medium properties. For some exotic materials the balance between density and shell stiffness is perfect for transparency in water, but, for most common metals some stiffness and mass need to be added. The proposed design of the internal oscillator is an axisymmetric distribution of lengthwise ribs (stiffeners) supporting a central rod. With $J=16$ stiffeners, the low frequency flexural resonance of the shell are suppressed. Simulations of an aluminum shell tuned to water with and acrylic oscillator show transparency up to $ka=0.8$.

A fluid saturated array of tuned elastic shells is capable of steering waves. The unit cell is comprised of the shell-spring-mass system surrounded by a fluid region. For the cell to have desired density and bulk modulus, the effective properties of the shell are tuned according to the filling fraction. A cylindrical-to-plane wave lens has been simulated with a 7x7 array of such cells. At frequencies $ka<0.7$ the shells behave as an effective medium. The acoustic energy from the monopole source at the center of the lens is steered away from the corners, decreasing the pressure by a factor of 3, to the faces, where the pressure increases by a factor of 7. This was achieved with the wave passing through only three rows of shells. 

\section{Acknowledgement}  \label{ack}
This work was supported by ONR though MURI Grant No. N00014-13-1-0631, and ULI Grant No. N00014-13-1-0417. Many thanks to Dr. Maria Medeiros of ONR (Code 333) for supporting and Dr. Stephen O'Regan of NSWCCD (Code 7220) for hosting the first author during this research. 

%%%%%%%%%%%%%%%%%%%%%%%%%%%%%%%%%%%%%%%%%%%%%%%%%%%%%%%%%

%%%%%%%%%%%%%%%%%%%%%
%\bibliographystyle{unsrt}%natbib}%unsrtnat}%doipubmed}%harvard}% plain}%uabbrvnat}%
%\bibliography{an_big_bib.bib}
%\end{document}
%%%%%%%%%%%%%%%%%%%%%%%%%%%%%%%%%%%%%%%%%%%%%%%%%%%%%%%%%

%\bibpunct{(}{)}{;}{a}{}{,}
%\bibliography{ref}
%\bibliographystyle{jasa}

\end{document}